\documentclass[aps,prd,floats,floatfix,amssymb,twocolumn,superscriptaddress,nofootinbib]{revtex4-2}

\usepackage{physics}
\usepackage{amsmath}
\usepackage{amsfonts,amssymb}
\usepackage{graphicx,graphics}
\usepackage{hyperref}
\usepackage{ulem}
\usepackage{color}
\usepackage{mathrsfs}
\usepackage{bbold}
\usepackage{cancel}
\usepackage{xcolor}

\begin{document}

\title{Unruh-deWitt detector in impulsive plane wave spacetimes}

\author{Jo\~ao P. M. Pitelli}
\email[]{pitelli@unicamp.br}
\affiliation{Departamento de Matem\'atica Aplicada, Universidade Estadual de Campinas,
13083-859 Campinas, S\~ao Paulo, Brazil}
\author{Ricardo A. Mosna}
\email[]{mosna@unicamp.br}
\affiliation{Departamento de Matem\'atica Aplicada, Universidade Estadual de Campinas,
13083-859 Campinas, S\~ao Paulo, Brazil}
\affiliation{Instituto de F\'isica Gleb Wataghin, Universidade Estadual de Campinas,
13083-859, Campinas,  S\~ao Paulo, Brazil}

\begin{abstract}
    
We investigate the response function of an inertial Unruh-deWitt detector in an impulsive plane wave spacetime. Through symmetry considerations applied to the Wightman function, we demonstrate that the response function remains invariant for any inertial detector, even for those experiencing a discontinuous lightcone coordinate shift after interacting with the shockwave. This implies that the vacuum state in an impulsive plane wave spacetime is preserved under the associated spacetime symmetries. Additionally, we confirm that the quantum imprint of the shockwave, as discussed in [J. High Energ. Phys. 2021, 54 (2021)], is not an artifact and exhibits a distinct characteristic form. We identify this form by defining a “renormalized” response function for an eternally inertial detector, with Minkowski spacetime as a reference.
\end{abstract}

\maketitle

\section{Introduction}

An impulsive pp-wave in General Relativity is an infinitely thin sandwich gravitational wave with a Dirac delta profile. These shockwaves were first studied by Penrose using a cut-and-paste method, where Minkowski spacetime is cut at a null hyperplane and the two halves are joined with a warp, characterizing the wave profile~\cite{penrose}. Aichelburg and Sexl provided a physical example of such a gravitational shockwave, presenting an exact solution to Einstein’s field equations with the expected delta-singular profile~\cite{Aichelburg}. By solving Einstein's equations for an ultrarelativistic particle with velocity $v \to 1$ and mass $m \to 0$ (while keeping the energy $p$ constant), they determined the gravitational field of a massless particle. They concluded that the metric remains Minkowskian before and after the passage of the particle, yielding the solution:
\begin{equation}
ds^2 = -du\, dv + f(\vec{x}) \delta(u-u_0) du^2 + \delta_{ij} dx^i dx^j,
\label{general metric}
\end{equation}
where $x = (t, z, \vec{x})$, $i,j = 1,2$, $f(\vec{x}) = -8p \log{(x^2 + y^2)^{1/2}}$, and $u = u_0$ describes the particle’s trajectory, with $u = t - z$ and $v = t + z$ as lightcone coordinates.

For purely gravitational or electromagnetic waves, the singular profile is simpler, with $f(\vec{x}) = h_{ij} x^i x^j$ in Eq.~(\ref{general metric}). The matrix $h_{ij}$ takes the form:
\begin{equation}
h_{ij} = \left(
\begin{array}{cc}
a & 0\\
0 & b
\end{array}
\right)
\label{H diag}
\end{equation}
with $h^{i}_{\phantom{i}i} = (a + b) = 0$ (vanishing Ricci tensor) for the gravitational wave and $ a =b$ (vanishing Weyl tensor) for the electromagnetic wave. For the general case considered here, the matrix $h_{ij}$ can also be expressed (without loss of generality) as in Eq.~(\ref{H diag}), and the null energy condition requires that $a + b \leq 0$. Additionally, Einstein's equations imply that the energy-momentum tensor is given by:
\begin{equation} 
 T_{uu} = -\delta(u)\frac{\nabla^2 f(\vec{x})}{16\pi G} = -\delta(u)\frac{(a + b)}{16\pi G},
\end{equation}
which shows that the energy density $\rho(\vec{x}) = -\frac{(a + b)}{16\pi G}$ is constant over the propagating singular plane.

The geodesic equation in impulsive pp-wave spacetimes involves ill-defined products of distributions~\cite{steinbauer,balasin,gray,cho}. Nevertheless, it is possible to calculate the trajectory of inertial detectors in a mathematically rigorous way~\cite{steinbauer}, and this approach aligns with the more intuitive methods used in Refs.~\cite{,balasin,cho,gray}. At any rate, it is found that an inertial detector experiences a time delay after crossing the shockwave:
\begin{equation}
\Delta v = \frac{1}{2}(ax_0^2 + by_0^2),
\label{shapiro delay}
\end{equation} 
and a change in velocity:
\begin{equation}
\left\{
\begin{array}{l}
\Delta \dot{x} = ax_0,\\
\Delta \dot{y} = by_0,
\end{array}
\right.
\end{equation} 
where $\vec{x}_0 = (x_0, y_0)$ is the point where the detector crosses the shockwave at $u = u_0$. Notably, a static detector at $(x_0, y_0) \neq (0, 0)$ acquires a non-zero transverse velocity after the passage of the shockwave, an effect known as the velocity memory effect~\cite{zhang}.

The study of quantum fields in impulsive pp-wave spacetimes presents challenges, particularly due to the absence of a Cauchy surface for the specification of initial data. This issue was first pointed out by Penrose in Ref.~\cite{penrose2}, where he observed that any candidate spacelike surface $\Sigma$ for a Cauchy surface would fail to intersect a particular null geodesic, as well as any null-like trajectory beyond that geodesic (alternatively, if $\Sigma$ were to intersect such a geodesic, it would do so more than once). Nonetheless, quantum effects in impulsive pp-wave spacetimes can still be studied~\cite{klimvcik,garriga}. Since the spacetime is flat before and after the shockwave, the Klein-Gordon equation can be solved using plane wave modes. Typically, in-modes generate mixed out-modes due to the singular behavior of the wave equation at the wavefront. The Bogoliubov coefficients (connecting the in and out modes) can be calculated by taking the surface $u = u_0$ as the  ``Cauchy surface'' (at least) for the null trajectories with $v = \text{constant}$, since the trajectory with $u = \text{constant}$ is unaffected by the shockwave~\cite{garriga}. It turns out that the Bogoliubov coefficient $\beta$, which mixes positive and negative frequency modes, is zero, indicating no particle production by the shockwave.

However, in a recent study~\cite{gray}, Gray et al. observed that the response function of an Unruh-deWitt detector in an impulsive plane wave spacetime, interacting with a scalar field, is non-trivial when compared to Minkowski spacetime. This suggests that a quantum imprint of the impulsive wave exists, even with the Bogoliubov coefficient $\beta = 0$. However, their analysis assumed: i) a specific trajectory for the detector at $x = y = 0$, where there is no deflection by the shockwave and no velocity memory effect, and ii) a finite-time interaction between the detector and the scalar field. These assumptions lead to further questions: i) Does the Shapiro time delay~\cite{shapiro} in Eq.~(\ref{shapiro delay}) and the sudden change in velocity after crossing the shockwave affect the response function? ii) Is the non-trivial response function merely an artifact of the finite-time interaction, or does it have a distinct  ``characteristic'' profile in a shockwave spacetime?

In this paper, we address these questions by considering an eternally inertial detector in a shockwave spacetime characterized by the metric:
\begin{equation}
ds^2 = -du\, dv + \delta(u)(ax^2 + by^2) du^2 + dx^2 + dy^2,
\label{metric}
\end{equation}
interacting with a quantum scalar field. Without loss of generality, we set $u_0 = 0$. This spacetime has five Killing vector fields~\cite{balasin2}:
\begin{equation}
\left\{
\begin{array}{l}
\xi_1 = \partial_v,\\
\xi_2 = 2x \partial_v + u \partial_x,\\
\xi_3 = 2y \partial_v + u \partial_y,\\
\xi_4 = \partial_x + a \Theta(u) \xi_2,\\
\xi_5 = \partial_y + b \Theta(u) \xi_3,
\end{array}
\right.
\label{symm}
\end{equation}
where $\Theta(u)$ is the Heaviside step function. We first establish that the Wightman function respects these symmetries, implying that the response function is identical for any inertial detector, even those intersecting the shockwave at $\vec{x} \neq (0, 0)$. Then, by examining the short-distance behavior of the Wightman function, we demonstrate that the response function can be ``renormalized'' to extract a finite profile. 

It is important to clarify what we mean by a ``renormalized'' response function in the context of this paper. Note that the response function $\mathcal{F}^{\textrm{Mink}}(\Omega)$ for an inertial observer in Minkowski spacetime is well known to be zero for $\Omega > 0$ (excitation) and infinite for $\Omega < 0$ (deexcitation). However, the transition rate $\dot{\mathcal{F}}^{\textrm{Mink}}(\Omega)$, representing the number of transitions per unit time in an ensemble of identical detectors, remains finite.

In our setup, the coordinate $u$ serves as an affine parameter, and the pull-back of the Wightman function to the detector's trajectory, $W(u,u')$, depends on more than just $\Delta u = u - u'$, since $\partial_u$ is not a Killing field (energy conservation is disrupted by the shockwave, which conceptually accounts for the quantum imprint discussed in the following sections). As a result, the transition rate alone is not meaningful; only the response function provides insights into the interaction between the detector and scalar field on the shockwave spacetime.

For an inertial detector, the Wightman function $W(u,u')$ matches $W^{\textrm{Mink}}(u,u')$ before ($u < 0$ and $u' < 0$) and after ($u > 0$ and $u' > 0$) the shockwave, where spacetime is essentially Minkowski. In the nontrivial region $(uu' < 0)$, where $W(u,u')$ differs significantly from its Minkowski counterpart [see Eq.~(\ref{wightman}) below], we define
\begin{equation}
W(u,u') = \left[ W(u,u') - W^{\textrm{Mink}}(u,u') \right] + W^{\textrm{Mink}}(u,u').
\label{comparing mink}
\end{equation}
Notably, the response function calculated using $W(u,u') - W^{\textrm{Mink}}(u,u')$, that is, the response function rescaled by the infinite result in Minkowski spacetime (when $a = b = 0$ in Eq.~(\ref{H diag})), remains finite, retaining only the residual information from the shockwave.

It is important to emphasize that this process does not constitute a fundamental renormalization of the Green function, as such a subtraction lacks meaning given the distinct spacetimes involved. Instead, we are simply comparing the response function on the impulsive wave spacetime to what would be expected in Minkowski spacetime, i.e., in the absence of the shockwave.

\section{Wightman Function}

We consider a massless scalar field satisfying the equation
\begin{equation}
\square \Psi = 0.
\end{equation}
Given that $\xi_1 = \partial_v$ is an exact symmetry of the spacetime with the metric given by Eq.~(\ref{metric}), we can separate variables in the form $\Psi(x) = e^{-i k_{-}v} \psi(u, \vec{x})$. The resulting equation for $\psi(u, \vec{x})$ becomes~\cite{klimvcik}
\begin{equation}
i \frac{\partial \psi}{\partial u} = -\left(\frac{\nabla^2}{4k_{-}} + f(\vec{x}) \delta(u - u_0)\right)\psi,
\label{schro}
\end{equation}
where $\nabla^2 = \frac{\partial^2}{\partial x^2} + \frac{\partial^2}{\partial y^2}$ and $f(\vec{x}) = ax^2 + by^2$. We consider the vacuum state $|0\rangle_{\textrm{in}}$ corresponding to the in-modes 
\begin{equation}
u^{\textrm{in}}_{k_{-}, \vec{k}}(x) = \frac{1}{(2k_-)^{1/2}(2\pi)^{3/2}} e^{-i (k_- v + k_+ u - \vec{k} \cdot \vec{x})}, \quad u < u_0,
\end{equation}
where $k_{\pm} = \frac{k_t \pm k_z}{2}$ and $\vec{k} = (k_x, k_y)$. These states evolve via Eq.~(\ref{schro}) to~\cite{gray}
\begin{widetext}
\begin{equation}
u^{\textrm{in}}_{k_{-}, \vec{k}}(x) = \frac{1}{(2k_-)^{1/2}(2\pi)^{3/2}} e^{-i (k_- v - \vec{k} \cdot \vec{x})} \int{\frac{d\vec{x}' d\vec{k}'}{(2\pi)^2} e^{i(\vec{k} - \vec{k}') \cdot (\vec{x}' - \vec{x})} e^{-i\frac{\vec{k}'^2}{4k_-}(u - u_0)} e^{ik_- \Theta(u - u_0) f(\vec{x})}}, \quad u > u_0.
\label{in}
\end{equation}
\end{widetext}
The out-modes, corresponding to the vacuum state $|0\rangle_{\textrm{out}}$, are characterized by
\begin{equation}
u^{\textrm{out}}_{k_{-}, \vec{k}}(x) = \frac{1}{(2k_-)^{1/2}(2\pi)^{3/2}} e^{-i (k_- v + k_+ u - \vec{k} \cdot \vec{x})}, \quad u > u_0.
\label{out}
\end{equation}
Given the modes defined by Eqs.~(\ref{in}) and (\ref{out}), we can compute the Bogoliubov coefficients between $u_{k_-, \vec{k}}^{\textrm{in}}$ and $u_{k_-, \vec{k}}^{\textrm{out}}$. In particular, we have
\begin{equation}
u_{k_-, \vec{k}}^{\textrm{in}} = \int{dl_- d\vec{l} \left(\alpha_{k_-, l_-; \vec{k}, \vec{l}} \, u_{l_-, \vec{l}}^{\textrm{out}} + \beta_{k_-, l_-; \vec{k}, \vec{l}} \, u_{l_-, \vec{l}}^{\textrm{out}\ast}\right)},
\end{equation}
with $\beta_{k_-, l_-; \vec{k}, \vec{l}} = -\left(u_{l_-, \vec{l}}^{\textrm{in}}, u_{k_-, \vec{k}}^{\textrm{out}}\right)$, where $\left(\cdot, \cdot\right)$ is the usual Klein-Gordon inner product. This calculation has been performed in two different ways, with the same result. In Ref.~\cite{compere}, the surface $\Sigma$ given by $t = 0$ was considered, and in Refs.~\cite{klimvcik, garriga}, the surface $u = u_0$ was used. In both cases, it was found that $\beta_{k_-, l_-; \vec{k}, \vec{l}} = 0$\footnote{This can be easily seen by considering the surface $u = u_0$ as in Refs.~\cite{klimvcik, garriga}. Since $u_{k_-, \vec{k}}^{\textrm{in}}$ and $u_{l_-, \vec{k}}^{\textrm{out}\ast}$ are proportional to $e^{-i k_- v}$ and $e^{i l_- v}$, respectively, the Klein-Gordon inner product will be proportional to $\delta(k_- + l_-) = 0$.} and that there is no production of particles after the passage of the shockwave.

The Wightman function is given by 
\begin{equation}
\begin{aligned}
W(x, x') &= \phantom{}_\textrm{in}\langle 0\left|\Psi(x)\Psi(x') \right|0\rangle_\textrm{in}\\
& = \int{u_{k_-, \vec{k}}^{\textrm{in}}(x) u_{k_-, \vec{k}}^{\textrm{in}\ast}(x') dk_- d\vec{k}}.
\end{aligned}
\end{equation}
This integral was calculated in Refs.~\cite{gray, cho}, resulting in
\begin{widetext}
\begin{equation}
\begin{aligned}
W_\epsilon(x, x') &= -\frac{1}{4\pi^2} \left\{1 - a \Delta \Theta_u \frac{(u - i \epsilon)(u' + i \epsilon)}{u - u' - i \epsilon}\right\}^{-1/2} \left\{1 - b \Delta \Theta_u \frac{(u - i \epsilon)(u' + i \epsilon)}{u - u' - i \epsilon}\right\}^{-1/2}\\
&\times \left\{(\Delta u - i \epsilon)(\Delta v + i \epsilon) - \Delta \vec{x}^2 - \frac{a \Delta \Theta_u \left[(u' + i \epsilon)x - (u - i \epsilon)x'\right]}{\Delta u - i \epsilon - a \Delta \Theta_u (u - u_0 - i \epsilon)(u' - u_0 + i \epsilon)}\right.\\
&\quad - \left. \frac{b \Delta \Theta_u \left[(u' + i \epsilon)y - (u - i \epsilon)y'\right]}{\Delta u - i \epsilon - b \Delta \Theta_u (u - i \epsilon)(u' + i \epsilon)}\right\}^{-1},
\end{aligned}
\label{wightman}
\end{equation}
\end{widetext}
where $\Delta \Theta_u = \Theta(u) - \Theta(u')$, $\Delta u = u - u'$, $\Delta v = v - v'$, and $\Delta \vec{x} = (x - x', y - y')$, with the $i \epsilon$ prescription, i.e., $t \to t - i \epsilon$. Notice that we recover the Minkowski Wightman function
\begin{equation}
W_\epsilon(x, x') = -\frac{1}{4\pi^2} \frac{1}{(\Delta u - i \epsilon)(\Delta v - i \epsilon) - \Delta \vec{x}^2},
\end{equation}
when $u, u' > u_0$ and when $u, u' < u_0$.

Clearly, the Wightman function given in Eq.~(\ref{wightman}) is invariant under $v \to v + \alpha$ and $v' \to v' + \alpha$, since $W(x, x')$ is a function of $\Delta v$. Similarly, by exponentiating $\xi_2 = 2x \partial_v + u \partial_x$, we obtain the following change of coordinates:
\begin{equation}
\left\{
\begin{array}{l}
\tilde{u} = u,\\
\tilde{v} = v + 2x \lambda + \lambda^2 u,\\
\tilde{x} = x + \lambda u,\\
\tilde{y} = y,
\end{array}
\right.
\end{equation}
where $\lambda$ is an arbitrary parameter. By direct substitution, one can easily check that $W(x, x') = W(\tilde{x}, \tilde{x}')$. One can also check that $W(x, x')$ is invariant under the change of coordinates
\begin{equation}
\left\{
\begin{array}{l}
\tilde{u} = u,\\
\tilde{v} = v + 2x a \Theta(u) \lambda + a \Theta(u) \lambda^2 \left[1 - u a \Theta(u)\right],\\
\tilde{x} = x + \lambda \left[1 - u \Theta(u)\right],\\
\tilde{y} = y,
\end{array}
\right.
\end{equation}
corresponding to the symmetry $\xi_4 = \partial_x + a \Theta(u) \xi_2$. In the same way, one can show that the two remaining symmetries in Eq.~(\ref{symm}) are symmetries of the Wightman function, indicating that the vacuum $\left|0\right.\rangle_{\textrm{in}}$ is invariant under the group of symmetries of the spacetime.

\section{Unruh-deWitt Detector}

We consider a point-like detector with two energy levels $\left|0\right.\rangle_{\textrm{d}}$ and $\left|\Omega\right.\rangle_{\textrm{d}}$ interacting with a scalar field propagating on an impulsive plane wave spacetime. The interaction Hamiltonian is given by
\begin{equation}
H_{\textrm{int}} = c\chi(\tau)m(\tau)\Psi(x(\tau)),
\end{equation}
where $c$ is a small coupling constant, $\chi(\tau)$ is the switching function, and $m(\tau) = e^{-i \tau \Omega} |\Omega \rangle_{\textrm{d}} \langle 0|_{\textrm{d}} + e^{i \tau \Omega} |0 \rangle_{\textrm{d}} \langle \Omega|_{\textrm{d}}$ is the idealized detector's monopole moment operator.

It is well known that the response function for the Unruh-deWitt detector in a general curved spacetime is given by~\cite{birrell}
\begin{equation}
\begin{aligned}
    \mathcal{F}(\Omega) = \lim_{\epsilon\to 0^{+}} &\int_{-\infty}^{\infty}\mathrm{d}\tau\int_{-\infty}^{\infty}\mathrm{d}\tau' e^{-i\Omega(\tau-\tau')} \\
    &\times \chi(\tau)\chi(\tau^{\prime})W_\epsilon(x(\tau), x(\tau^{\prime})).
\end{aligned}
\label{response function}
\end{equation}
By introducing a change of coordinates in the form $w := \tau$, $s := \tau - \tau'$ when $\tau' < \tau$, and $w := \tau'$, $s := \tau' - \tau$ when $\tau < \tau'$, we arrive at~\cite{schlicht}
\begin{equation}
\begin{aligned}
    \mathcal{F}(\Omega) = 2\lim_{\epsilon\to 0^{+}} &\int_{-\infty}^{\infty}\mathrm{d}w \int_{0}^{\infty}\mathrm{d}s \,\chi(w)\chi(w-s) \\
    &\times \textrm{Re}\left[e^{-i \Omega s}W_\epsilon(w, w-s)\right].
\end{aligned}
\label{double}
\end{equation}
If we consider a switching function of the form
\begin{equation}
\chi(\tau) = \Theta(\tau + T)\Theta(T - \tau),
\end{equation}
i.e., the detector interacts with the quantum field from $\tau = -T$ to $\tau = T$, the double integral in Eq.~(\ref{double}) can be viewed as the limit when $T \to \infty$ of the following integral:
\begin{equation}
\begin{aligned}
 I &= \int_{-T}^{T}\mathrm{d}w \int_{0}^{w+T}\mathrm{d}s\,\textrm{Re}\left[e^{-i \Omega s}W_\epsilon(w, w-s)\right].
\end{aligned}
\end{equation}
After changing the order of integration, $I$ is equivalent to
\begin{widetext}
\begin{equation}
I = \int_{0}^{T}\mathrm{d}s \int_{s-T}^{T}\mathrm{d}w\,\textrm{Re}\left[e^{-i \Omega s}W_\epsilon(w, w-s)\right] + \int_{T}^{2T}\mathrm{d}s \int_{s-T}^{T}\mathrm{d}w\,\textrm{Re}\left[e^{-i \Omega s}W_\epsilon(w, w-s)\right],
\end{equation}
\end{widetext}
with the last integral going to zero in the limit $T \to \infty$.\footnote{In fact, there should be a switching function $\chi(w)\chi(w-s)$ inside the integral $\int_T^{2T}{\mathrm{d}s \int_{s-T}^{T}\mathrm{d}w}$. This switching function can be chosen as a function in $C_0^{\infty}(I)$, where $I \subset \mathbb{R}$ is a closed interval, or as a function that decays sufficiently fast at infinity. In this case, it is easy to see that $ \lim_{T\to\infty}{\int_T^{2T}{\mathrm{d}w}} = 0$.} We then arrive at
\begin{equation}
\begin{aligned}
    \mathcal{F}(\Omega) = 2\lim_{\epsilon\to 0^{+}} &\int_{0}^{\infty}\mathrm{d}s \int_{-\infty}^{\infty}\mathrm{d}w \,\textrm{Re}\left[e^{-i \Omega s}W_\epsilon(w, w-s)\right].
\end{aligned}
\end{equation}

In the next section, we will parametrize the geodesics using the parameter $u$ (which will serve as an affine parameter). The Wightman function is exactly the Minkowski Wightman function $W^{\textrm{Mink}}$ for $u, u' < 0$, which corresponds to $-\infty < w < 0$, and for $u, u' > 0$, which corresponds to $w > s$. This shows that
\begin{equation}
\begin{aligned}
    \mathcal{F}(\Omega) &= 2\lim_{\epsilon\to 0^{+}} \int_{0}^{\infty}\mathrm{d}s \int_{-\infty}^{0}\mathrm{d}w \,\textrm{Re}\left[e^{-i \Omega s}W^{\textrm{Mink}}_\epsilon(w, w-s)\right] \\
    &\quad + 2\lim_{\epsilon\to 0^{+}} \int_{0}^{\infty}\mathrm{d}s \int_{s}^{\infty}\mathrm{d}w \,\textrm{Re}\left[e^{-i \Omega s}W^{\textrm{Mink}}_\epsilon(w, w-s)\right] \\
    &\quad + 2\lim_{\epsilon\to 0^{+}} \int_{0}^{\infty}\mathrm{d}s \int_{0}^{s}\mathrm{d}w \,\textrm{Re}\left[e^{-i \Omega s}W_\epsilon(w, w-s)\right].
\end{aligned}
\end{equation}

If we consider an eternally switched on detector, i.e., $\chi(\tau) = 1$, the first two integrals in the above equation diverge for $\Omega < 0$. This is not surprising, as the response function for an eternally inertial detector in Minkowski spacetime also diverges for de-excitation (note that $\mathcal{F}^{\textrm{Mink}}(\Omega) = 0$ for $\Omega > 0$). This divergence is related to the fact that $W^{\textrm{Mink}}$ is invariant under time translation. However, if we define a "renormalized" response function as
\begin{equation}
\begin{aligned}
 \mathcal{F}_\textrm{ren}(\Omega) &= \mathcal{F}(\Omega) - \mathcal{F}^{\textrm{Mink}}(\Omega) \\
 & = 2\lim_{\epsilon\to 0^{+}} \int_{0}^{\infty}\mathrm{d}s \int_{0}^{s}\mathrm{d}w \,\chi(w)\chi(w-s) \\
 &\quad \times \textrm{Re}\left\{e^{-i \Omega s}\left[W_\epsilon(w, w-s) - W^{\textrm{Mink}}(w, w-s)\right]\right\},
\end{aligned}
\label{renormalized}
\end{equation}
we may obtain a finite result,  and a possibly non-zero response for $\Omega > 0$. 

\section{Detector's Trajectory}

In this section, we demonstrate that the response function for a general inertial detector in a shockwave spacetime can be determined by examining the simplest case: a static observer at $x = y = 0$. To achieve this, we present the trajectory for an inertial detector as derived in Refs.~\cite{steinbauer,gray,balasin,cho}, while deferring a more detailed discussion of the derivation to the appendix. There, we briefly address the limitations of geodesic equations in the distributional sense and how to handle this issue using two approaches: the naive approach from Refs.~\cite{gray,balasin,cho} and the more formal approach outlined in Ref.~\cite{steinbauer}.

  An inertial detector in a shockwave spacetime follows the trajectory given by


\begin{widetext}
\begin{equation}
\left\{
\begin{array}{l}
u = u,\\
v = v_0 + u + (\dot{x}_0^2 + \dot{y}_0^2)u + \Theta(u)(a x_0^2 + b y_0^2) + u\Theta(u)\left[(a^2 x_0^2 + b^2 y_0^2) +2 (a x_0 \dot{x}_0 + b y_0 \dot{y}_0)\right],\\
x = x_0 + \dot{x}_0 u + u\Theta(u)(a x_0),\\
y = y_0 + \dot{y}_0 u + u\Theta(u)(b y_0),
\end{array}
\right.
\label{geod final}
\end{equation}
\end{widetext}
where $\displaystyle x_0 = \lim_{u \to 0^{-}}{x(u)}$, $\displaystyle \dot{x}_0 = \lim_{u \to 0^{-}}{\dot{x}(u)}$, $\displaystyle y_0 = \lim_{u \to 0^{-}}{y(u)}$, and $\displaystyle \dot{y}_0 = \lim_{u \to 0^{-}}{\dot{y}(u)}$.

Consider an inertial detector with $x_0 = y_0 = 0$, $\dot{x}_0 = \dot{y}_0 = 0$, and $v_0 = 0$. This detector is not deflected by the passage of the shockwave. After a change of coordinates of the form $u \to u$, $v \to v + v_0$, $x \to x$, and $y \to y$ (using the Killing field $\xi_1 = \partial_v$), we arrive at
\begin{equation}
\left\{
\begin{array}{l}
u = u,\\
v = v_0 + u,\\
x = 0,\\
y = 0,
\end{array}
\right.
\end{equation}

Next, by applying the transformation $u \to u$, $v \to v + 2x \dot{x}_0 + \dot{x}_0^2 u$, $x \to x + \dot{x}_0 u$, and $y \to y$ (exploring the Killing field $\xi_2$ in Eq.~(\ref{symm})), we obtain
\begin{equation}
\left\{
\begin{array}{l}
u = u,\\
v = v_0 + u + \dot{x}_0^2 u,\\
x = \dot{x}_0 u,\\
y = 0,
\end{array}
\right.
\end{equation}
Then, by performing the coordinate change $u \to u$, $v \to v + 2 x a \Theta(u) x_0 + a \Theta(u) x_0^2(1 + u a)$, $x \to x + x_0[1 + au \Theta(u)]$, $y \to y$ (using the Killing field $\xi_4$ in Eq.~(\ref{symm})), we have
\begin{equation}
\left\{
\begin{array}{l}
u = u,\\
v = v_0 + u + \dot{x}_0^2 u + \Theta(u)(a x_0^2) + u \Theta(u)(2a^2 x_0^2 + a x_0 \dot{x}_0),\\
x = x_0 + \dot{x}_0 u + u \Theta(u)(a x_0),\\
y = 0,
\end{array}
\right.
\end{equation}
Finally, we apply the same steps using the symmetries $\xi_3$ and $\xi_5$ to obtain Eq.~(\ref{geod final}).

At every step, we know that the Wightman function remains invariant. Therefore, to analyze the response function around any inertial trajectory, we can place the detector along the trajectory
\begin{equation}
\left\{
\begin{array}{l}
u = u,\\
v = u,\\
x = 0,\\
y = 0,
\end{array}
\right.
\label{simple trajectory}
\end{equation}
This will significantly simplify our calculations.

\section{Response Function of an Eternally Inertial Unruh-deWitt Detector}

The Wightman function is relatively simple along the trajectory given in Eq.~(\ref{simple trajectory}). It is expressed as
\begin{equation}
\begin{aligned}
&W_\epsilon(u, u') = -\frac{1}{4\pi^2} \frac{1}{(u - u' - i \epsilon)} \\
&\times \frac{1}{\sqrt{u - u' - a \Delta\Theta_u u u' - i(\epsilon - a (u - u') \epsilon)}} \\
&\times \frac{1}{\sqrt{u - u' - b \Delta\Theta_u u u' - i(\epsilon - b (u - u') \epsilon)}}.
\end{aligned}
\end{equation}

We then have
\begin{widetext}
\begin{equation}
\mathcal{F}_\textrm{ren}(\Omega) = 2\textrm{Re}\left[\int_0^{\infty} \mathrm{d}s e^{-i\Omega s} \int_0^{s} \mathrm{d}w \left(-\frac{1}{4\pi^2} \frac{1}{(s - i\epsilon)} \frac{1}{\sqrt{s - a w (w - s) - i (1 - a s)\epsilon}} \frac{1}{\sqrt{s - b w (w - s) - i (1 - b s)\epsilon}} + \frac{1}{4\pi s^2}\right)\right],
\label{integral}
\end{equation}
\end{widetext}
where \( W^{\textrm{Mink}}(w, w-s) = -\frac{1}{4\pi s^2} \).

The inner integral in Eq.~(\ref{integral}) can be treated analytically using Refs.~\cite{grads, handbook}. The exterior integral can then be evaluated numerically using the  software  Mathematica~\cite{mathematica}. We will work out the two cases of main interest in detail: \( b = -a \) (gravitational wave) and \( b = a \) (electromagnetic wave). The response function for other cases, with \( a + b \leq 0 \) due to the null energy condition, behaves similarly.

\subsection{Electromagnetic Wave}
Let us first consider the case \( a = b = -\lambda \), with \( \lambda > 0 \). We have
\begin{equation}
\begin{aligned}
W(w, w-s) &= -\frac{1}{4\pi^2 (s - i \epsilon)} \\
&\times \frac{1/\lambda}{w^2 - s w + \frac{s}{\lambda} + i \epsilon \left(s - \frac{1}{a}\right)}.
\end{aligned}
\end{equation}
For \( \Delta = s^2 - 4\left(\frac{s}{\lambda}\right) < 0 \), i.e., when \( s < 4/\lambda \), the equation \( w^2 - s w + \frac{s}{\lambda} \) has no real roots, and the term \( i\epsilon \) becomes irrelevant in the limit \( \epsilon \to 0^+ \). Then we have (see Eq.~2.172 of Ref.~\cite{grads})
\begin{equation}
\int_0^s \mathrm{d}w W(w, w-s) = -\frac{4}{4 \pi^2 \lambda s} \arctan\left(\frac{s}{\sqrt{-\Delta}}\right).
\label{fs1}
\end{equation}
For \( \Delta > 0 \), i.e., when \( s > 4/\lambda \), we have
\begin{equation}
\begin{aligned}
W_\epsilon(w, w-s) &= -\frac{1}{4\pi^2 (s - i \epsilon)} \left(\frac{1}{\lambda \sqrt{\Delta}}\right) \\
&\times \left[\frac{1}{w - w_+ - i \epsilon} - \frac{1}{w - w_- + i \epsilon}\right],
\end{aligned}
\end{equation}
with \( w_\pm = \frac{1}{2}\left(s \pm \sqrt{\Delta}\right) \). Since \( s = w - w' \), we see that \( w_{\pm} \) are solutions of the equation
\begin{equation}
\frac{1}{w} - \frac{1}{w'} = \lambda.
\end{equation}
This is analogous to a lens in geometrical optics with focal length \( 1/\lambda \) and is related to the fact that every geodesic with perpendicular incidence will focus to a point at \( u = 1/\lambda \)~\cite{cho}.

Note that the integrals \( \int_0^s \frac{\mathrm{d}w}{w - w_{\pm}} \) are undefined since \( w_{\pm} < s \) and \( \frac{1}{w - w_{\pm}} \) are not integrable near \( w_{\pm} \). Therefore, we use the Sokhotski-Plemelj formula:
\begin{equation}
\lim_{\epsilon\to 0^+} \frac{1}{x \pm i\epsilon} = \mathcal{P} \frac{1}{x} \mp i \pi \delta(x)
\end{equation}
so that
\begin{equation}
\begin{aligned}
f(s) \equiv &\lim_{\epsilon\to 0^+} \int_0^s \mathrm{d}w\, W_\epsilon(w, w-s) \\
= &-\frac{1}{4\pi^2 s} \left(\frac{1}{\lambda \sqrt{\Delta}}\right) \\
&\times \left\{\log\left[\frac{\left(\frac{\frac{s}{2} - \frac{\sqrt{\Delta }}{2}}{\frac{s}{2} + \frac{\sqrt{\Delta }}{2}}\right)}{\left(\frac{\frac{s}{2} + \frac{\sqrt{\Delta }}{2}}{\frac{s}{2} - \frac{\sqrt{\Delta }}{2}}\right)}\right] - 2\pi i\right\},
\end{aligned}
\label{fs}
\end{equation}
where
\begin{equation}
\begin{aligned}
&\int_0^s \mathrm{d}w\, \mathcal{P}\frac{1}{w - w_{\pm}} = \lim_{\eta\to 0^+}
\Bigg\{\int_0^{w_{\pm} - \eta} \mathrm{d}w\, \frac{1}{w - w_{\pm}} \\
&+ \int_0^{w_{\pm} + \eta} \mathrm{d}w\, \frac{1}{w - w_{\pm}}\Bigg\}.
\end{aligned}
\end{equation}
The remaining integral, namely,
\begin{equation}
2 \textrm{Re}\left[\int_0^\infty f(s) \mathrm{d}s\right],
\end{equation}
with \( f(s) \) given by Eqs.~(\ref{fs1}) and~(\ref{fs}), can then be solved numerically.

\begin{figure}[htb!]
\includegraphics[scale=0.35]{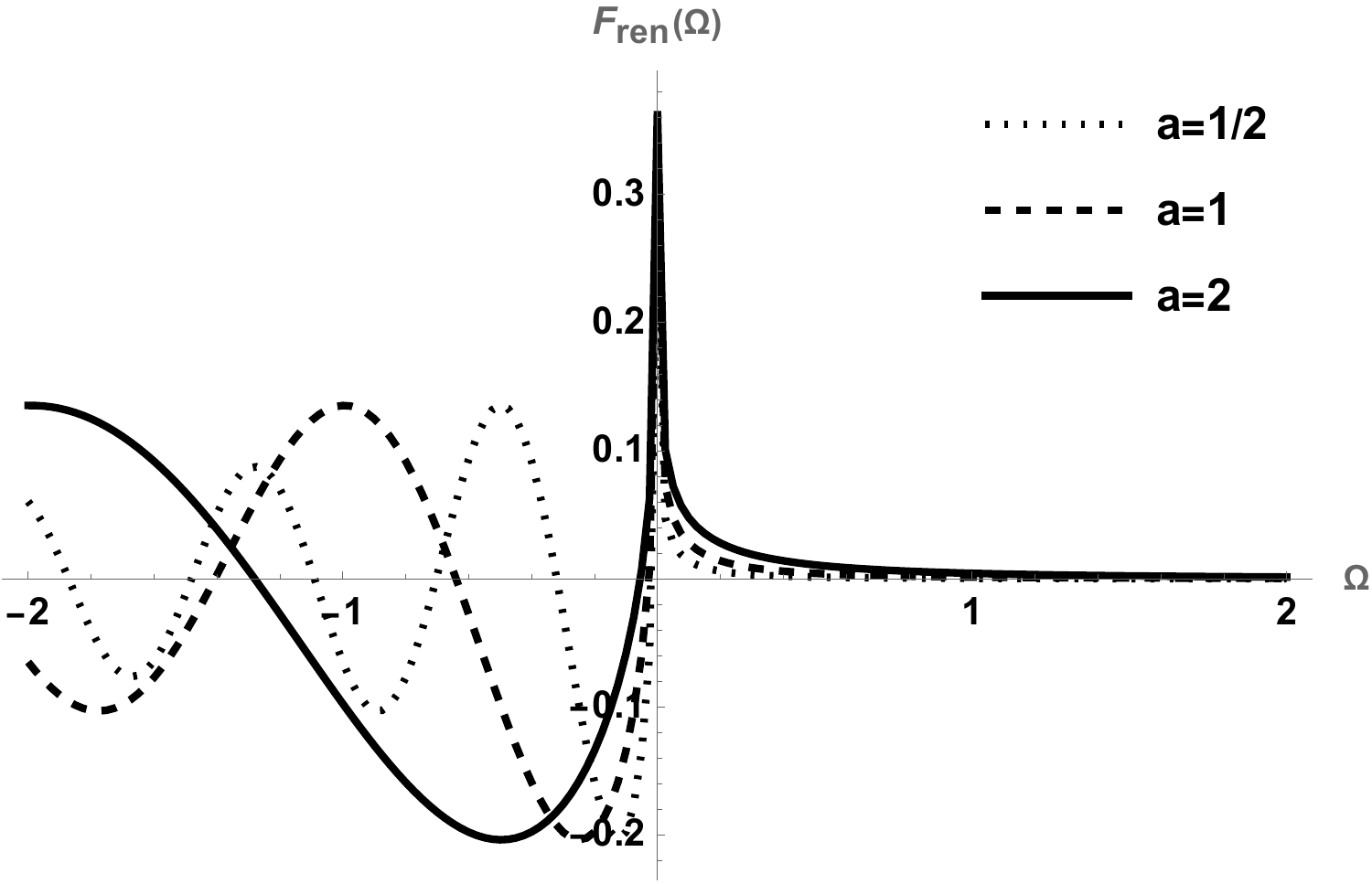}
\caption{Quantum imprint for the electromagnetic case for \( a = \frac{1}{2}, 1, 2 \). Note that the response function for \( \Omega > 0 \) grows with the energy parameter \( a \). For \( \Omega < 0 \), \( \mathcal{F}_\textrm{ren}(\Omega) \) oscillates more slowly as \( a \) increases.} 
\label{fig1}
\end{figure}

The resulting \( \mathcal{F}_\textrm{ren}(\Omega) \) is shown in Fig.~\ref{fig1}. Notice that \( \mathcal{F}_\textrm{ren}(\Omega) \) decays quickly as \( \Omega \to \infty \) and oscillates with a slow decay as \( \Omega \to -\infty \). This characteristic behavior can be identified with the quantum imprint discussed in Ref.~\cite{gray}.
 We note that the parameter $a$ defines an energy scale for the problem, which determines the range of $\Omega > 0$ where $\mathcal{F}_\textrm{ren}(\Omega)$ remains significant. For $\Omega < 0$, the period of oscillation increases with this energy scale, reflecting the impact of the parameter $a$ on the behavior of the response function.

\subsection{Gravitational Wave}

\begin{figure}[htb!]
\includegraphics[scale=0.35]{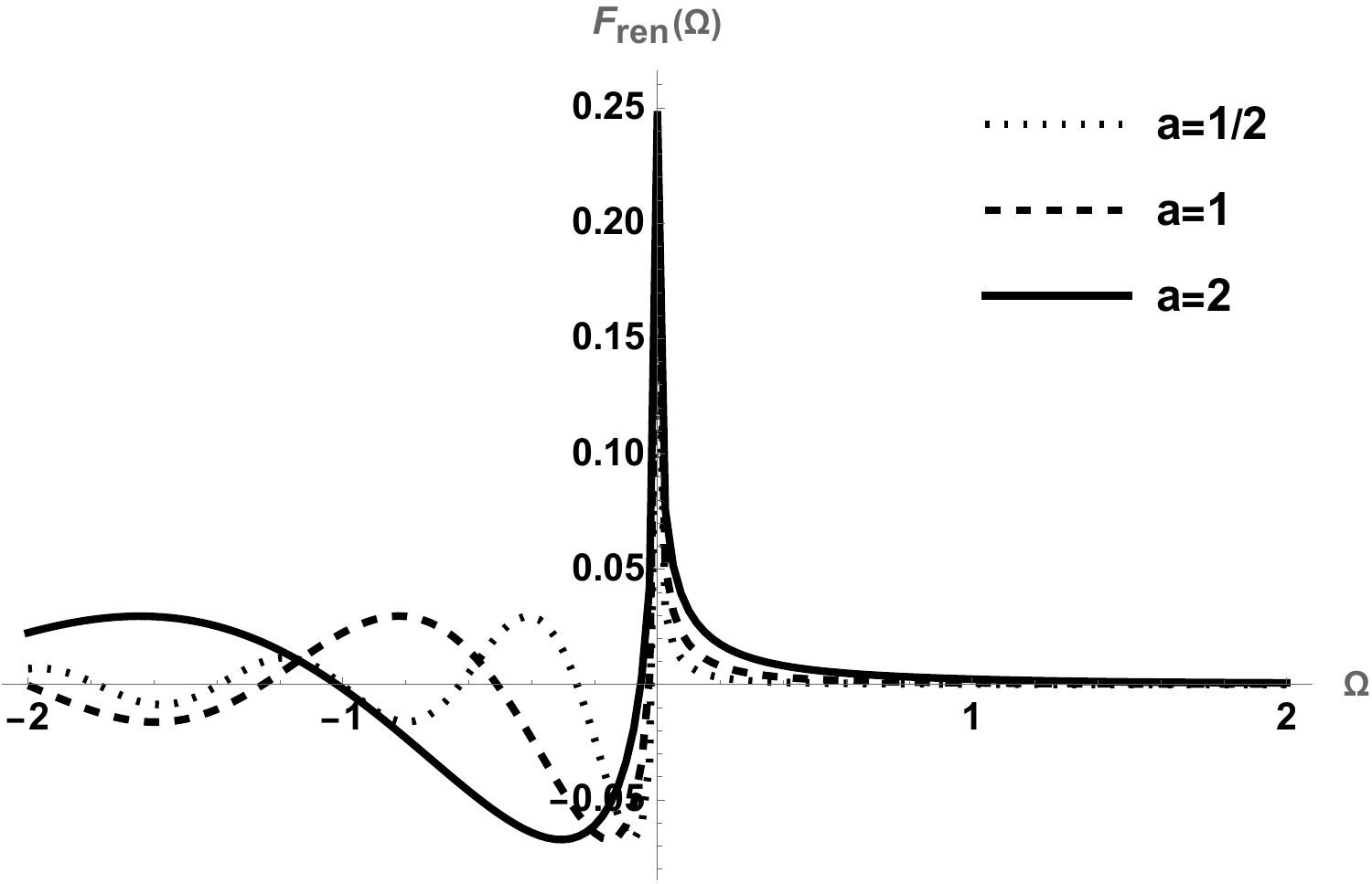}
\caption{Quantum imprint for the gravitational case for \( a = \frac{1}{2}, 1, 2 \).} 
\label{fig2}
\end{figure}

Now we consider the case \( a = -b = \lambda > 0 \). We have
\begin{equation}
W(w, w-s) = -\frac{1}{4\pi^2 s} \frac{1/\lambda}{\sqrt{u^4 - 2 s u^3 + s^2 u^2 - \frac{s^2}{a^2}}},
\end{equation}
where we discarded the \( i\epsilon \) term since the resulting term is integrable. The polynomial \( u^4 - 2 s u^3 + s^2 u^2 - \frac{s^2}{a^2} \) has two real roots: \( w_1 = \frac{s}{2} - \frac{1}{2}\sqrt{s^2 + \frac{4s}{a}} < 0 \) and \( w_4 = \frac{s}{2} + \frac{1}{2}\sqrt{s^2 + \frac{4s}{\lambda}} > s \) for all \( s > 0 \). The other two roots, \( 0 < w_3 = \frac{s}{2} - \frac{1}{2}\sqrt{s^2 - \frac{4s}{\lambda}} < w_4 = \frac{s}{2} + \frac{1}{2}\sqrt{s^2 - \frac{4s}{\lambda}} < s \), are real only for \( s > 4/\lambda \). Using Ref.~\cite{handbook}, we have, for \( 0 < s < 4/\lambda \),
\begin{equation}
\int_0^s W(w, w-s) \mathrm{d}w = -\frac{1}{4\pi^2 \lambda s} g \left(F(\phi_2, k^2) - F(\phi_1, k^2)\right),
\end{equation}
where \( F \) is the incomplete elliptic integral of the first kind, and
\begin{equation}
\begin{aligned}
&k = \sqrt{\frac{(w_4 - w_1)^2 + (A - B)^2}{4 A B}},\\
&A = \sqrt{(w_1 - \beta)^2 + \alpha^2},\\
&B = \sqrt{(w_4 - \beta)^2 + \alpha^2},\\
&g = \frac{1}{\sqrt{AB}},\\
&\alpha^2 = -\frac{(w_2 - w_3)^2}{4},\\
&\beta = \frac{(w_2 + w_3)}{2},\\
&\phi_2 = \arccos\left[\frac{(w_4 - s)B - (s - w_1)A}{(w_4 - s)B + (s - w_1)A}\right],\\
&\phi_1 = \arccos\left[\frac{w_4 B + w_1 A}{w_4 B - w_1 A}\right].
\end{aligned}
\end{equation}
For \( s > 4/\lambda \), we have
\begin{equation}
\begin{aligned}
\int_0^s W(w, w-s) \mathrm{d}w &= -\frac{1}{4\pi^2 \lambda s} g \Big[F(\phi_{1}, k_1^2) - F(\phi_{2}, k_2^2) \\
&\quad + \frac{1}{i}F(\phi_{3}, k_3^2) - \frac{1}{i}F(\phi_{4}, k^2)\Big],
\end{aligned}
\end{equation}
where
\begin{equation}
\begin{aligned}
&g = \frac{2}{\sqrt{(w_4 - w_2)(w_3 - w_1)}},\\
&k_1 = k_2 = \sqrt{\frac{(w_4 - w_3) (w_2 - w_1)}{(w_4 - w_2 )(w_3 - w_1)}},\\
&\phi_1 = \pi/2,\\
&\phi_2 = \arcsin\left(\sqrt{\frac{(w_4 - w_3)(0 - w_1)}{(w_2 - w_1)(w_4 - 0)}}\right),\\
&k_2 = \sqrt{\frac{(w_2 - w_3) (w_4 - w_1)}{(w_4 - w_3 )(w_2 - w_1)}},\\
&\phi_3 = \pi/2,\\
&k_3 = \sqrt{\frac{(w_3 - w_2) (w_4 - w_1)}{(w_4 - w_2) (w_3 - w_1)}},\\
&\phi_4 = \arcsin\left(\sqrt{\frac{(w_4 - w_2) (s - w_3)}{(w_4 - w_3) (s - w_2)}}\right),\\
&k_4 = \sqrt{\frac{(w_4 - w_3) (w_2 - w_1)}{(w_4 - w_2) (w_3 - w_1)}}.
\end{aligned}
\end{equation}
The resulting \( \mathcal{F}_\textrm{ren}(\Omega) \) is shown in Fig.~\ref{fig2}. Notice that \( \mathcal{F}_\textrm{ren}(\Omega) \) once again decays rapidly as \( \Omega \to \infty \) and oscillates with a slower decay as \( \Omega \to -\infty \). 
 Notice that, in the gravitational case, the energy density along the propagating null hyperplane is zero. However, the energy scale $a$ remains paramount, once again determining the range of $\Omega > 0$ for which $\mathcal{F}_{\textrm{ren}}(\Omega)$ is appreciable.

\section{Conclusions}
In this paper, we have undertaken a detailed analysis of the quantum imprint left by an impulsive plane wave on an eternally inertial Unruh-deWitt detector. By exploring the response function of the detector interacting with a massless scalar field, we have found the exact form for the quantum imprint.  Despite the divergence of the exact response function, we have demonstrated that a finite renormalized response function can be defined, effectively capturing the quantum signature of the shockwave.

We have examined two specific cases, namely the gravitational and electromagnetic waves, to illustrate the behaviour of the quantum imprint under different conditions. Our results show that, in both cases, there is  an oscillatory behaviour with a slow decay  for $\Omega<0$ (de-excitation) and a quick decay as $\Omega>0$ (excitation). 

An interesting question arises as to whether the subtraction defined in Eq.~(\ref{renormalized}) remains effective for non-inertial detectors interacting with potentially massive scalar fields, given that our renormalized response function is intrinsically dependent on the detector's trajectory. We believe this question warrants further investigation.
 
 In conclusion, the methods and results presented here provide a robust framework for investigating the quantum signatures of impulsive gravitational waves, with potential applications in both theoretical and observational contexts.

\acknowledgments

J. P. M. P. thanks the support provided in part by Conselho Nacional de Desenvolvimento Cient\'{i}fico e Tecnol\'{o}gico (CNPq, Brazil), Grant No. 311443/2021-4.
R.A.M. was partially supported by Conselho Nacional de Desenvolvimento Cient\'{\i}fico e Tecnol\'{o}gico (CNPq, Brazil) under Grant No. 316780/2023-5.
We are indebted to Professors G. Matsas  and J. Garriga for clarifying several questions during the development of this work. J. P. M. P.  also thanks  the Department of Quantum Physics and Astrophysics of the Universitat de Barcelona (where part of this work was developed) for the kind hospitality.

\appendix*
\section{}

In this appendix, we briefly discuss two different ways to derive Eq.~(\ref{geod final}), closely following Refs.~\cite{steinbauer,balasin}.

Given the metric in Eq.~(\ref{metric}), the geodesics in this spacetime can be found by solving the following system of equations (for details, see Refs.~\cite{balasin,gray,cho}):
\begin{widetext}
\begin{align}
\frac{d^2 u}{d\tau^2} &= 0\label{eqgeod1},\\
\frac{d^2 v}{d\tau^2} &= \delta'(u)(a x^2 + b y^2)\left(\frac{du}{d\tau}\right)^2 + 4\frac{du}{d\tau} \delta(u)\left(a x \frac{dx}{d\tau} + b y \frac{dy}{d\tau}\right) \label{eqgeod2},\\
\frac{d^2 x}{d\tau^2} &= a x \delta(u)\left(\frac{du}{d\tau}\right)^2\label{eqgeod3},\\
\frac{d^2 y}{d\tau^2} &= b y \delta(u)\left(\frac{du}{d\tau}\right)^2\label{eqgeod4},
\end{align}
\end{widetext}
where $\tau$ is the proper time. The first equation implies that $u$ serves as an affine parameter, so we use $u$ in place of $\tau$. Thus, Eq.~(\ref{eqgeod3}) becomes
\begin{equation}
\ddot{x}(u) = a x(u) \delta(u),
\label{eq3becomes}
\end{equation}
indicating that $x(u)$ must be a straight line before and after the shockwave. Assume that $x(u) = x_0 + \dot{x}_0 u$ for $u < 0$ and $x(u) = A + Bu$ for $u > 0$, so that
\begin{equation}
x(u) = x_0 + A \Theta(u) + \left[\dot{x}_0 + B \Theta(u)\right] u.
\label{soleq3}
\end{equation}
Using the identities
\begin{align}
\frac{d}{du}\Theta(u) &= \delta(u),\label{id1}\\
\frac{d}{du}\left[u \Theta(u)\right] &= \Theta(u),\label{id2}
\end{align}
and applying $u \delta(u) = 0$ in Eq.~(\ref{id2}), we obtain
\begin{equation}
\frac{d^2 x}{du^2} = A \delta'(u) + B \delta(u).
\end{equation}
By substituting Eq.~(\ref{soleq3}) into Eq.~(\ref{eq3becomes}), we find
\begin{equation}
\frac{d^2 x}{du^2} = a \left\{ x_0 + A \Theta(u) + \left[\dot{x}_0 + B \Theta(u)\right] u \right\} \delta(u).
\end{equation}
Thus, $A = 0$ and $B = a x_0$, giving
\begin{equation}
x(u) = x_0 + \dot{x}_0 u + a x_0 u \Theta(u).
\label{x solution}
\end{equation}
Similarly, we obtain
\begin{equation}
y(u) = y_0 + \dot{y}_0 u + b y_0 u \Theta(u).
\label{y solution}
\end{equation}

The second-order equation for $v(u)$ in Eq.~(\ref{eqgeod2}) is more subtle, as it involves ill-defined distributions like $\Theta(u) \delta(u)$. Consequently, the right side of Eq.~(\ref{eqgeod2}) is not well-defined in the distributional sense.

However, there are two ways to handle this issue. In the naive approach (Refs.~\cite{balasin,cho,gray}), applying the norm constraint
\begin{equation}
g_{\mu\nu}\frac{dx^\mu}{du} \frac{dx^\nu}{du} = -1,
\end{equation}
yields the following equation for $v(u)$:
\begin{equation}
\frac{dv}{du} = 1 + (a x^2 + b y^2) \delta(u) + \left(\frac{dx^i}{du}\right)^2.
\end{equation}  
By substituting Eqs.~(\ref{soleq3}) and~(\ref{y solution}) and using $\Theta^2(u) = \Theta(u)$, we obtain
\begin{equation}
\begin{aligned}
\frac{dv}{du} &= 1 + (a x_0^2 + b y_0^2) \delta(u) + \dot{x}_0^2 + \dot{y}_0^2\\
&\quad + \left[ 2 (a x_0 \dot{x}_0 + b y_0 \dot{y}_0) + a^2 x_0^2 + b^2 y_0^2 \right] \Theta(u),
\end{aligned}
\end{equation}
which is meaningful in the distributional sense and can be integrated to yield
\begin{equation}
\begin{aligned}
v &= v_0 + u + (\dot{x}_0^2 + \dot{y}_0^2) u + \Theta(u) (a x_0^2 + b y_0^2) \\ 
&\quad + u \Theta(u) \left[ a^2 x_0^2 + b^2 y_0^2 + 2 (a x_0 \dot{x}_0 + b y_0 \dot{y}_0) \right].
\end{aligned}
\label{sol v}
\end{equation}
Taking the second derivative of this solution gives
\begin{equation}
\begin{aligned}
\frac{d^2 v}{du^2} &= (a x_0^2 + b y_0^2) \delta'(u)\\
&\quad + \delta(u) \left[ 2 (a x_0 \dot{x}_0 + b y_0 \dot{y}_0) + a^2 x_0^2 + b^2 y_0^2 \right].
\end{aligned}
\label{eq v first}
\end{equation}
However, inserting solutions~(\ref{x solution}) and~(\ref{y solution}) into the right side of Eq.~(\ref{eqgeod2}) yields
\begin{equation}
\begin{aligned}
\frac{d^2 v}{du^2} &= (a x_0^2 + b y_0^2) \delta'(u)\\
&\quad + 2 \delta(u) (a x_0 \dot{x}_0 + b y_0 \dot{y}_0)\\
&\quad + 2 (a^2 x_0^2 + b^2 y_0^2) \delta(u) \Theta(u),
\end{aligned}
\label{eq v second}
\end{equation}
where we used $u \delta'(u) = -\delta(u)$ and $u^2 \delta(u) = 0$. Clearly, Eqs.~(\ref{eq v first}) and~(\ref{eq v second}) differ. This discrepancy arises because the second-order equation for $v(u)$ is distributionally ill-defined. However, it easy to see that the two can be reconciled by choosing $\Theta(0) = \frac{1}{2}$. 
In contrast, Steinbauer in Ref.~\cite{steinbauer} derives these solutions using a more rigorous approach (where the condition $\Theta(0) = 1/2$ is not explicitly imposed). Here, the delta functions in Eqs.~(\ref{eqgeod1}) to~(\ref{eqgeod4}) are regularized with a mollifier $\rho_\epsilon(x) = \frac{1}{\epsilon} \rho(x / \epsilon)$, where $\rho$ has compact support on $[-1, 1]$ and $\int_{-1}^{1} \rho(x) \, dx = 1$. First, the delta function is represented as
\begin{equation}
\delta_\epsilon(x) = \int_{-\infty}^{\infty} dy \, \delta(y) \frac{1}{\epsilon} \rho\left(\frac{x - y}{\epsilon}\right) = \frac{1}{\epsilon} \rho\left(\frac{x}{\epsilon}\right) := \rho_\epsilon(x),
\label{delta reg}
\end{equation}
so that the relevant geodesic equations become
\begin{equation}
\begin{aligned}
\ddot{v}_\epsilon(u) &= f(x_\epsilon^i(u)) \dot{\rho}_\epsilon(u) + 2 \partial_i f(x_\epsilon^i(u)) \dot{x}_\epsilon(u) \rho_\epsilon(u), \\
\ddot{x}^i_\epsilon(u) &= \frac{1}{2} \partial_i f(x_\epsilon^i(u)) \rho_\epsilon(u),
\end{aligned}
\label{system nonlinear}
\end{equation}
where $f(x^i) = a x^2 + b y^2$. This nonlinear system~(\ref{system nonlinear}) has a unique solution for small $\epsilon$.

In this formalism, the relation $\Theta \delta = \frac{1}{2} \delta$ can be justified by noting that the delta and Heaviside functions both originate from the same source: the $\delta$-shaped wave profile. Thus, both factors in the ill-defined product $\delta \Theta$ involve the same regularization, leading to
\begin{equation}
\Theta_\epsilon(x) = \int_{-\infty}^{\infty} dy \, \Theta(y) \frac{1}{\epsilon} \rho\left(\frac{x - y}{\epsilon}\right) = \frac{1}{\epsilon} \int_{-\infty}^{x} dy \, \rho\left(\frac{y}{\epsilon}\right)
\end{equation}
and also Eq.~(\ref{delta reg}). Thus,
\begin{equation}
\begin{aligned}
(\Theta \delta, \varphi) &= \int_{-\infty}^{\infty} dx \frac{1}{\epsilon} \rho\left(\frac{x}{\epsilon}\right) \int_{-\infty}^{x} dy \frac{1}{\epsilon} \rho\left(\frac{y}{\epsilon}\right) \varphi(x) \\
&= \int_{-\infty}^{\infty} \int_{-\infty}^{x} dx \, dy \, \rho(x) \rho(y) \varphi(\epsilon x) \\
&= C \varphi(0) + \mathcal{O}(\epsilon),
\end{aligned}
\label{order epsilon}
\end{equation}
where $\varphi(x)$ is a test function and
\begin{equation}
\begin{aligned}
C &= \int_{-\infty}^{\infty} \int_{-\infty}^{x}  \, \rho(x) \rho(y)dy \, dx \\
&= \frac{1}{2} \int_{-\infty}^{\infty} \int_{-\infty}^{\infty} \, \rho(x) \rho(y) dy \, dx \\
&= \frac{1}{2} \left( \int_{-\infty}^{\infty} \rho(x) \, dx \right)^2 = \frac{1}{2}.
\end{aligned}
\end{equation}


\begin{thebibliography}{99}

\bibitem{penrose}
R. Penrose, {\it  Twistor quantization and curved space-time}, Int. J. Theor. Phys. {\bf 1}, 61, 1968.

\bibitem{Aichelburg}
P.C. Aichelburg and R.U. Sexl, {\it On the gravitational field of a massless particle}, Gen. Rel.
Grav. {\bf 2}, 303, 1971.



\bibitem{steinbauer}
R. Steinbauer, {\it Geodesics and geodesic deviation for impulsive gravitational waves}, 	J. Math. Phys. { \bf 39} 2201, 1998.
 
 
\bibitem{balasin}
H. Balasin, {\it Geodesics for impulsive gravitational waves and the multiplication of distributions}, Class. Quant. Grav. {\bf 14}, 455, 1997.
 


\bibitem{gray}
F. Gray, D. Kubiznak, T. May, S. Timmerman and E. Tjoa, {\it Quantum imprints of gravitational shockwaves}, J. High Energ. Phys. {\bf 54}, 2021.

\bibitem{cho}
H. T. Cho, {\it Wightman function and stochastic gravity noise kernel in impulsive plane wave spacetimes
}, Phys. Rev. D {\bf 108}, 105007, 2003.



\bibitem{zhang}
P.-M. Zhang, C. Duval and P. A. Horvathy, {\it Memory Effect for Impulsive Gravitational Waves}, Class. Quantum Grav. {\bf 35}, 065011, 2018.

\bibitem{penrose2}
R. Penrose, {\it A remarkable property of plane waves in general relativity}, Rev. Mod. Phys. {\bf 37}, 215, 1965.

\bibitem{klimvcik}
C. Klim\v c\'ik,{\it  Quantum field theory in gravitational shock wave background}, Phys. Lett. B {\bf 208}, 373, 1988.

\bibitem{garriga}
J. Garriga and E. Verdaguer, {\it Scattering of quantum particles by gravitational plane waves}, Phys. Rev. D {\bf 43}, 391, 1991.

\bibitem{shapiro}
I. I. Shapiro, {\it  Fourth test of general relativity}, Phys. Rev. Lett. {\bf 13}, 789, 1964.

\bibitem{balasin2}
P. C. Aichelburg and H. Balasin, {\it Generalized symmetries of impulsive gravitational waves }, Class. Quant. Grav. {\bf 14}, A31, 1997.

\bibitem{compere}
G. Comp\`ere, J. Long and M. Riegler, {\it Invariance of Unruh and Hawking radiation under matter-induced supertranslations}, JHEP {\bf 05}, 053, 2019.

 
\bibitem{birrell}
N. D. Birrell and P. C. W.  Davies, {\it Quantum Fields in Curved Space}. Cambridge University Press, Cambridge, 1982.

\bibitem{schlicht}
S. Schlicht, {\it Considerations on the Unruh effect: causality and regularization},
Class. Quantum Grav. {\bf 21}, 4647 (2004).
\bibitem{mathematica}
Wolfram Research, Inc., {\it Mathematica, Version 14.0}, Champaign, IL (2024).

\bibitem{grads}
I. S. Gradshteyn and I. M. Ryzhik, {\it Table of integrals, series, and products},
Academic Press, New York (1980).

\bibitem{handbook}
P. F. Byrd and  M. D. Friedman, {\it Handbook of elliptic integrals for
engineers and physicists}, Springer-Verlag, Berlin (1954). 
\end{thebibliography}
\end{document}